\documentclass[prl,twocolumn,showpacs,amsmath,amssymb,floatfix,superscriptaddress]{revtex4}
\usepackage{graphicx}
\usepackage{times}

\begin{document}

\title{Phonon Diodes and Transistors from Magneto-acoustics}

\author{Sophia R. Sklan}
\affiliation{Department of Physics, Massachusetts Institute of Technology, 
Cambridge, Massachusetts 02139, USA}
\author{Jeffrey C. Grossman}
\affiliation{Department of Materials Science and Engineering,
Massachusetts Institute of Technology, Cambridge, MA 02139, USA.}


\begin{abstract}
By sculpting the magnetic field applied to magneto-acoustic materials,
phonons can be used for information processing.
Using a combination of analytic and numerical techniques, we demonstrate
designs for diodes (isolators) and transistors that are independent of their
conventional, electronic formulation.
We analyze the experimental feasibility of these systems, including the
sensitivity of the circuits to likely systematic and random errors.
\end{abstract}


\maketitle

Heat is ubiquitous.
It accompanies almost any form of energy loss in real systems,
but is one of the most difficult phenomena to control precisely.
The most successful utilizations of heat (e.g. heat engines
and heat pumps) essentially treat it as a homogenous current.
However, consider crystals, where heat is
often transported by electrons, photons, and phonons.
There exist impressive arrays of devices for controlling both
electrons and photons (down to specific modes and locations),
but no equivalent toolkit for phonons.
Perhaps the starkest example of this is computing,
where strict control of a signal's state is compulsory and commonplace.
Recent efforts have sought to extend this degree of control
to phonons, to realize devices like diodes, transistors, logic,
and memory \cite{Phononics,Phn Logic,Phn Diode,Phn Transistor,Phn Mem}.
Throughout this process the assumption that all computers should
be the strict analog of electronic computers has been implicit. Since
information in electronics is scalar (high or low voltage = 1 or 0),
it has been assumed that information from phonons would be encoded
in temperature (hot or cold = 1 or 0).
Similarly, since electronics uses pn junctions
for constructing circuits, interface effects have been considered for phonon diodes.
Hence, research has thus far focused on nano-structures \cite{CNT Rect,VO Rect}
or 1D materials \cite{Phononics,Phn Logic,Phn Diode,Therm Rect}, where interface effects
are strong, but fabrication was difficult.

\begin{figure}[h] \begin{center}
\includegraphics[scale=1]{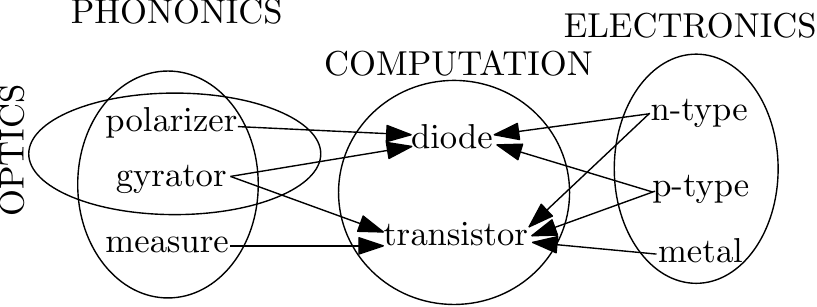}
\caption{\label{fig:compare}Constitutive construction of logic elements in
electronics, optics, and phononics. Circles represent classes of signals: electronic,
optical, phononic, and logical. Elementary devices for controlling these signals are
listed in each circle. For electronics, optics, and phononics, the basic elements are
typically a single material or interface. Logical signals
are an abstraction, so devices are defined by their effect on signals.
Arrows indicate which basic elements are required to construct these logic elements for
a given signal.}
\end{center}
\end{figure}

Abandoning the assumption that phononic and electronic computing are
strictly analogous presents a host of new opportunities.
Here, we make an analogy to optical computing.
We encode information in the polarization of a phonon current 
(transverse vertical or horizontal = 1 or 0).
Our operators therefore modify some generic elliptic polarization,
i.e. gyrators (which rotate the polarization angle) and polarizers (which project the polarization) from which we can construct
diodes and transistors. The relationship between devices used in electronics,
optics, and phononics and the abstract logic elements is shown in Figure \ref{fig:compare}
\footnote{In optics isolators are essentially diodes \cite{Pht Isolator}.}.
To make these, we require systems that break time-reversal, rather than
reflection, symmetry $-$ that is, we require a magnetic field.
For the magnetic field to have a measurable effect upon the phonon current,
we focus on magneto-acoustic (MA) materials.
These materials were first described by Kittel, who noted that they could be
used to create ``gyrators, isolators [diodes], and other nonreciprocal acoustic elements" \cite{Kittel},
but subsequent research on MA focused on other applications
(e.g. acoustic control of magnetization) \cite{GG Book,AFM Th,TGG Exp, Luthi Book, Magnons}.
MA coupling is a bulk effect found in commercially available
materials, so fabrication is easier compared to the nano-structures of the electronic analogy.

In this article, we employ a combination of analytic and numerical methods to demonstrate that phononic logic elements (diodes and transistors) can be designed outside of an electronic computing paradigm. 
Our results confirm that MA polarizers and gyrators, when combined with a means of generating and measuring phonon currents, are sufficient to realize logic elements.
Further, we show how present experimental techniques are likely sufficient to actualize logic elements that are reliably insensitive to errors.
Taken together, these results reveal the potential for an under-explored class of phonon logic gates.


The goal of this work is to explore the feasibility of frequency-dependent phonon computing.
In order to tackle this, knowledge of the phonon dispersion's dependence on fixed (e.g. length)
and tunable (e.g. magnetic field) parameters is necessary.
Hence, we begin with the dispersion relation for two special geometries.

When the magnetic field is oriented along the length of the MA
($\vec{H}\Vert\vec{k}$, where $H$ is the applied field and $k$ the phonon
wavevector), we have the circular birefringence (acoustic Faraday effect (AFE)) necessary
for a gyrator \cite{GG Book}.
The dispersion relation is:
\begin{equation}
k_\pm(\omega) = \frac{T}{l}(1-\frac{A}{B\pm T\mp i})^{-1/2},
\end{equation}
where $k_\pm$ are the right and left circularly polarized wavevectors (at fixed
frequency $\omega$), $l$ the natural length scale $\tau \sqrt{c_{1313}/\rho}$ (
where $\tau$ the shortest relevant lifetime, typically the magnon lifetime, $c_{1313}$
the stiffness constant, and $\rho$ the density), $T$ the dimensionless frequency 
$\omega\tau$, $B$ the dimensionless field strength $\gamma\tau H$ ($\gamma$ is
the gyroscopic ratio), and $A$ the dimensionless coupling constant 
$\gamma b_2^2 \tau/c_{1313}M_0$ ($b_2$ is the MA constant and $M_0$
the saturation magnetization of the MA, assuming a net ferromagnetic
moment exists).
Conversely, when the magnetic field is oriented perpedicular to the length of the
MA ($\vec{H}\perp\vec{k}$), we have linear birefringence 
(Cotton-Mouton effect), necessary for a polarizer \cite{GG Book}.
The dispersion relation for the mode polarized along the magnetic field is:
\begin{equation}
k_\Vert(\omega) = \frac{T}{l}(1-\frac{AB}{(1-iT)^2+B(B+4\pi\gamma\tau M_0)})^{-1/2},
\end{equation}
whereas the mode polarized perpendicular to the magnetic field is unaffected by the magnetic
field ($k_\perp(\omega)=T/l=\omega\sqrt{\rho/c_{1313}}$).
In both cases, there will be both real and imaginary components to the dispersion,
corresponding to birefringence and dichroism \footnote{Dichroism can
have any sign in MA, as energy can be added or extracted from the magnetic field}.

\begin{figure*}[!ht] \begin{center}
\includegraphics[scale=0.75]{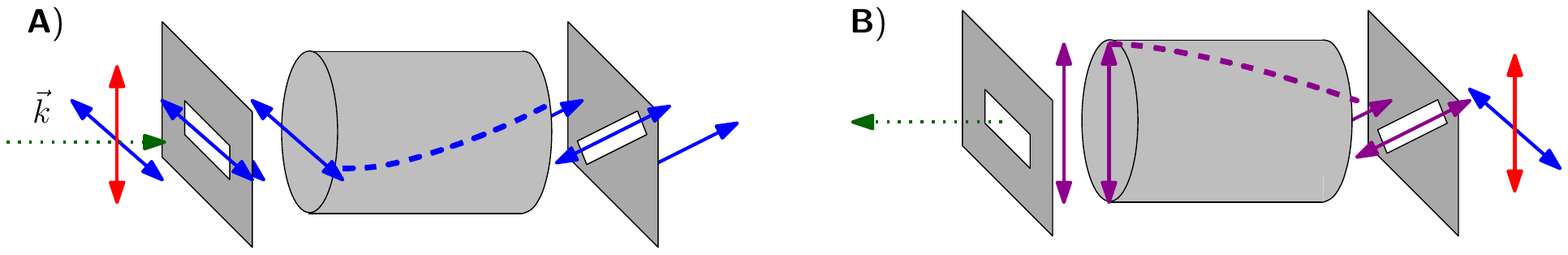}
\includegraphics[scale=0.2]{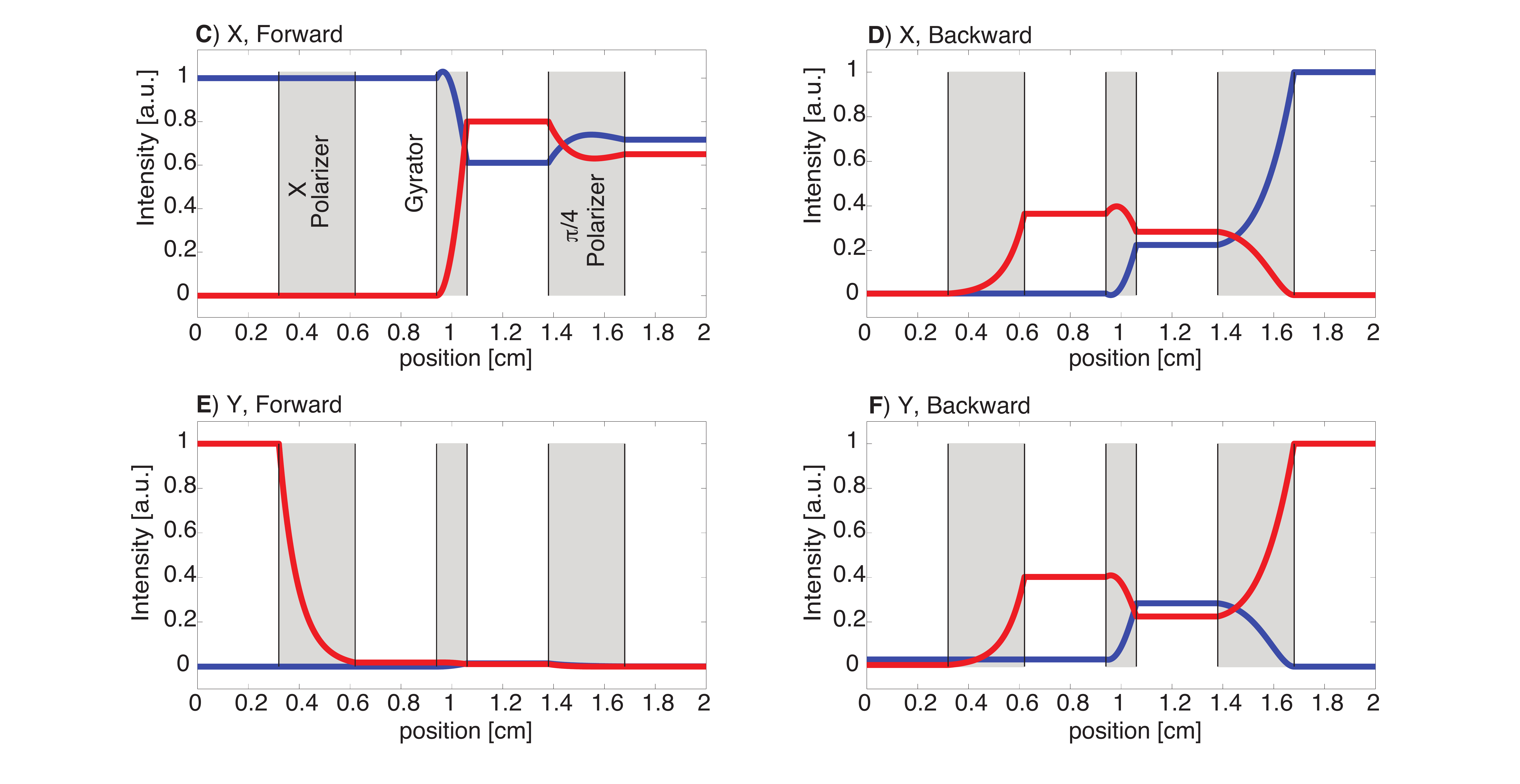}
\caption{\label{fig:diode}(A), (B): Schematic diagram for andiode.
Constructed of a gyrator (cylinder) between by two polarizers (rectangles,
 gaps indicate the polarization that is allowed to pass).
Blue and red lines indicate the x and y polarizations, with purple ((B) only) being a superposition
of both.
Green lines denote the direction of signal propogation(the wavevector, $\vec{k}$).
(A) Forward operation.
Unpolarized signal enters, becomes polarized, gyrated, and then leaves.
(B) Backward operation.
Unpolarized signal enters, becomes polarized, gyrated, and then blocked.
(C)-(F): Operation of a diode. Blue and red lines indicate polarized intensities. Grey rectangles indicate MA.
(C) and (D) have the
incoming signal x-polarized (allowed polarization), whereas (E) and (F) have the
incoming signal y-polarized. (C) and (E) have the signal approaching from the front of the
diode, while D) and (F) have the signal approaching from the back.
Only in (C) is the signal transmitted.}
\end{center}
\end{figure*}

In optics, diodes are constructed by sandwiching a $\pi/4$ gyrator between
two linear polarizers (oriented by $\pi/4$ with respect
to each other (see Fig. \ref{fig:diode})) \cite{Pht Isolator}.
A signal entering in the forward mode, passes through the first polarizer,
acquires a rotated polarization from the gyrator, and emerges polarized along 
the second polarizer.
Conversely, a signal in the reverse direction is polarized and then
acquires the same rotation in polarization, emerging orthogonal to the second
polarizer.
Both polarizers and gyrators can be constructed from MA by tuning the magnetic field.
For a diode, one must select magnetic field strengths (at fixed frequency) that yield weak dichroism
for the gyrator (even weak circular dichroism can prevent complete destructive interference,
as we see below) and strong dichroism for the polarizer.

Our independent parameters for designing the gyrators and polarizers are field strength,
phonon frequency, and MA length.
We assume that the properties of the MA are fixed, taking values from
representative experiments \cite{EXP}.
The Phonon frequency is selected to be 10 GHz (slightly larger than in \cite{EXP}).
$k(H)$ is then calculated for each dispersion and used to select reasonable magnetic fields that give desirable
ratios of birefringe to dichroism (0.01T for the polarizers and 0.1T for the gyrator).
Lastly, lengths are selected such that the gyration ($\theta=L(k_+^\prime-k_-^\prime)/2$
) and filtering ($\alpha=\mathrm{exp}(-k_\Vert^{\prime\prime}L)$) are effective.
The resulting circuit is then modeled by numerically evaluating the phase acquired by the phonon
current at each stage.
The results are plotted in Figure \ref{fig:diode}, where we find the circuit successfully blocks ($>95\%$ loss of intensity)
all signals except the desired polarization and direction.
Because the AFE's solutions have opposite signs in their imaginary components,
the amplification found in the forward mode is expected (the polarizers suppress it
in the reverse).

\begin{figure}[h] \begin{center}
\includegraphics[scale=0.6]{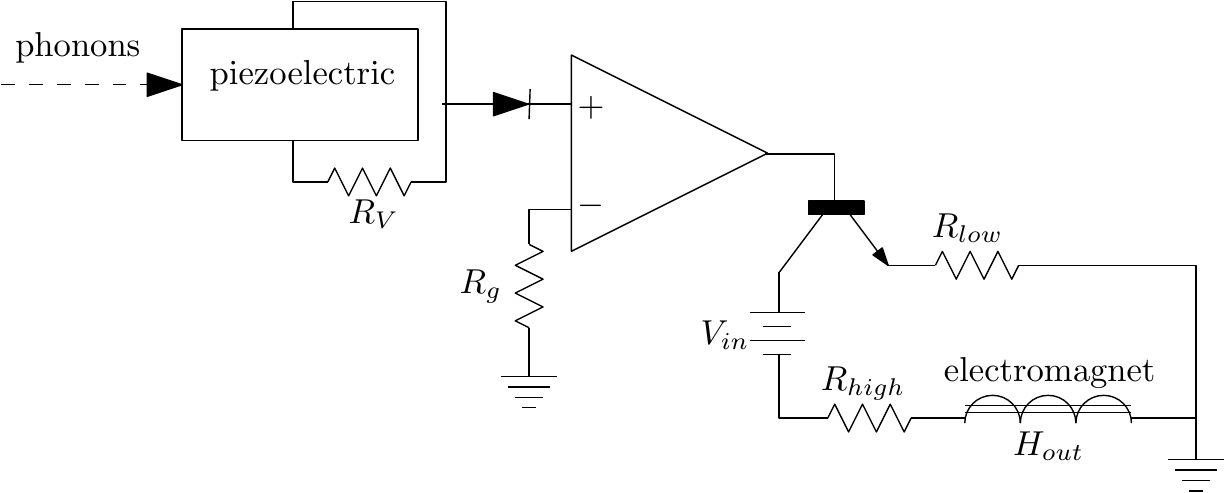}
\caption{\label{fig:measurement}Measurement operator.
Phonon current passes through a piezoelectric, transducing an electronic signal proportional
to the polarized phonon amplitude.
This voltage is then rectified and amplified (via a diode and an op amp).
The resulting voltage is used to switch between driving a current through an electromagnet
(producing a magnetic field) or not ($R_{high}\gg R_{low}$).
Operation shown here is for a magnetic field withheld when a signal is detected.}
\end{center}
\end{figure}

Turning to the transistor, we require a more complicated approach.
Firstly our transistor requires a measurement apparatus.
This type of measurement remains difficult, but we show a heuristic approach in Fig. \ref{fig:measurement}.
While there may be more efficient experimental realizations,
the form presented here benefits from its conceptual simplicity. 
The different stages of detection and transduction (piezoelectric), rectification (electronic diode),
amplification (op amp), and application (electronic transistor, electromagnet) are all differentiated
and are in principle realizable.

\begin{figure*}[ht!] \begin{center}
\includegraphics[scale=0.75]{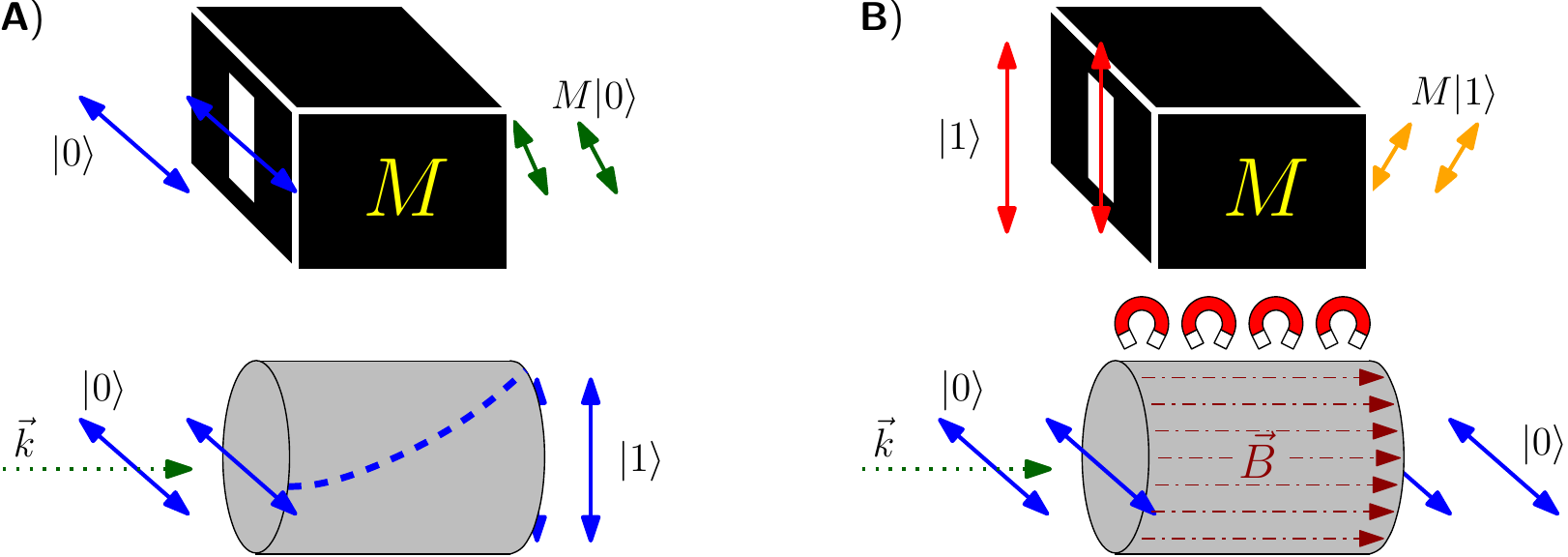}
\includegraphics[scale=0.3]{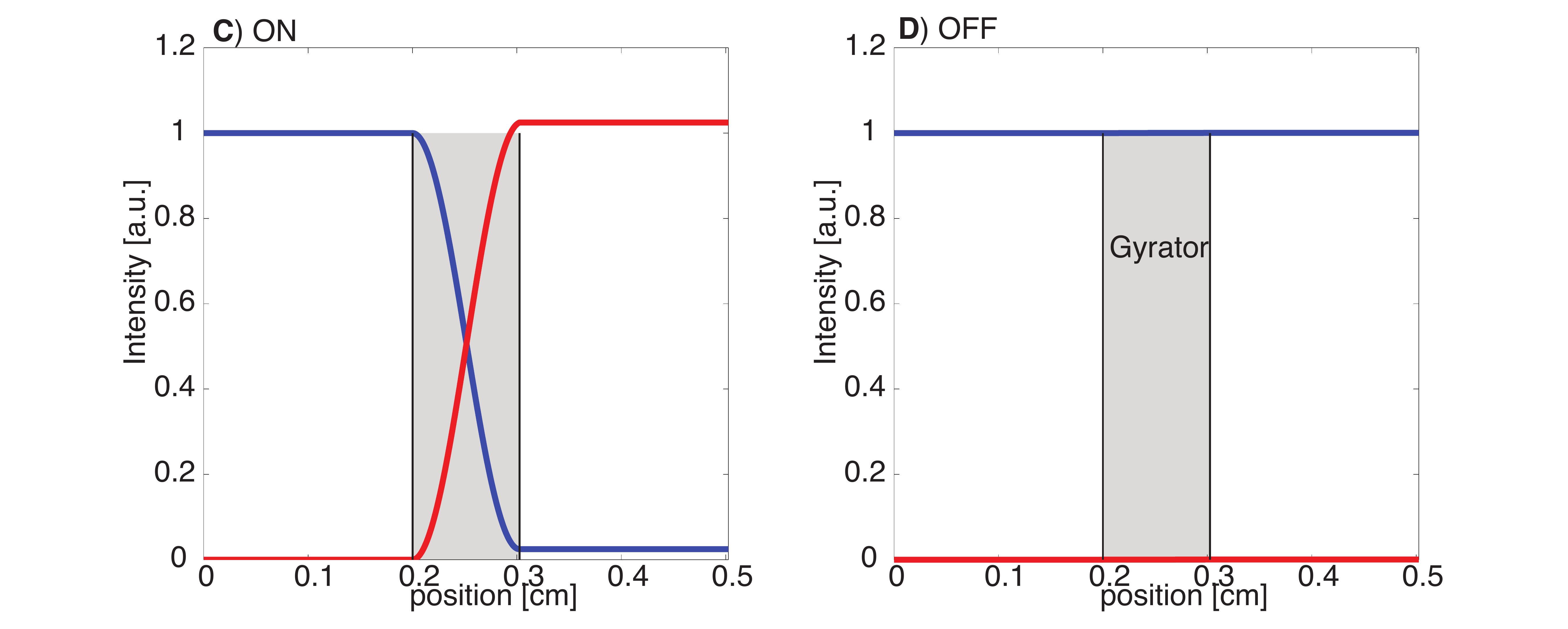}
\caption{\label{fig:transistor}Operation of a transistor. Same conventions as Figure \ref{fig:diode}.
A) schematic of the transistor in the on state.
The black box marked M denotes the measument operator, as sketched in Figure \ref{fig:measurement}.
B)  schematic of the transistor in the off state.
The magnets denote the source of the magnetic field (dark red lines) applied to the gyrator.
C) transistor is on (no suppressing field), switching 0 to 1.
D) transistor is off (suppressing field applied), no switching occurs.}
\end{center}
\end{figure*}

Given a measurement operator, we send a fixed logical 0 signal into a gyrator, and then determine if a magnetic
field should be applied by measuring the amplitude of the phonon current for one polarization.
If this polarization exceeds some threshold, a magnetic field is supressed (the gyration
is strongest as $B\to0$ 
\footnote{The model used here does not include thermal
fluctuations reducing the MA's magnetization at remanence that of saturation.
A small, non-zero field is likely prefereable.};
remanence magnetization provides the necessary magnetic field to keep the gyrator working).
Conversely, when the threshhold is not reached, then a magnetic field will be applied, suppressing
the birefringence and partially cancelling the gyration (perfect cancellation requires $B \to \infty$).
These two operations are summarized in Fig \ref{fig:transistor}A,B.

For the transistor to work as a logic operation, the gyration should be $\pi/2$.
Using the same process as in the design of the diode (magnetic field of $10^{-4}$T for off and 0.5T for on),
we model the transistor in Figure \ref{fig:transistor}C,D.
In doing so we abstract the measurement device, focusing instead on the effect of
applying or suppressing a magnetic field.

We find that there are imperfections in each operating regime.
When the gyrator is off (field applied), the relatively modest size of the field implies a small
gyration is still present.
Whereas, when the gyrator is on, circular dichroism prevents perfect cancellation of the
left and right circularly polarized modes, resulting in a small horizontally polarized remnant.
For the specific case of an incoming signal at $\theta_{in}=0$ and the length optimized
for $\theta_{out}=\pi/2$, the outgoing polarization angle is limited by
\begin{equation}
\tan\theta = \coth \frac{\pi}{2} \frac{\mathrm{Im}[k_+-k_-]}{\mathrm{Re}[k_+-k_-]}.
\end{equation}
\footnote{Numerical calculations slightly exceed this limit, since we find numerically that the
maximum $\theta$ occurs for a slightly thinner transistor than would be predicted by
$L=\pi/(k_+^\prime-k_-^\prime)$.
This gain is not large enough to merit abandoning the proposed limit.}
While this can be accounted for by allowing some fuzziness to the range of polarizations that are
deemed logical 0 or 1 (indeed, the piezoelectric transduction in Fig \ref{fig:measurement} is relatively insensitive
to the undesired polarization), there is a more stringent limit implied by these errors.
Since the undesireable gyrations in the off state will accumulate, there exists a
maximum total length of transistors that can be chained in series while maintaining well-separated logic states.
This problem can be surmounted in practice by applying a repeater circuit (which maps a noisy input
to a desired, less noisy value, as occurs in our transistor design when the signal is sent to the gate, not the source).
If we think of each gyrator in a series as tied to a separate gate input, then this also limits the number of independent
inputs in a logic operation that can be performed without using a repeater.
We can exceed this limit because multiple phonon currents can superimpose, but
practical difficulties in distinguishing between different inputs for super-imposed signals
make it unlikely that this distinction will do more than double the number of logical inputs.
To estimate the practical implications of this limitation, we consider the following encoding.
Logical 0 is [0,$\pi$/5] and logical 1 is [3$\pi$/10,$\pi$/2] (other quadrants mapping to the 1st by
reflection symmetry).
In this case, using our previous independent parameter values, we find that the number of (fixed length) gyrators goes as
\begin{equation}
N = \mathrm{floor}[6.4H^2-0.059|H|-0.0047],
\end{equation}
where $N$ is the maximum number of gyrators and $H$ the applied field strength in Tesla.
The minimum allowable field strength for the off state is therefore 0.4T.
While a similar limit for the on state exists, the insistence on $B\approx0$ for this regime makes it a weaker constraint
on the number of stages and field. 

The presence of circular dichroism in the AFE produces a systematic error
that limits computational power.
In addition to systematic errors, random errors can also corrupt a circuit's operation (be it diode or transistor).
While sufficiently thick polarizers are relatively insensitive to such errors (the damping is exponential),
gyrators can be quite sensitive.
In general, this sensitivity depends upon frequency and field strength.
To assess the sensitivity for an arbitrary case, we use the linearized equation of 
uncertainty propagation.
Expressing the result in fractional uncertainties gives:
\begin{equation}
|\frac{\sigma_{\Delta kL}}{\Delta kL}|^2 =(\frac{\sigma_L}{L})^2+|\frac{\partial \Delta k}{\partial H}|^2(\frac{\sigma_H}{H})^2+|\frac{\partial \Delta k}{\partial \omega}|^2(\frac{\sigma_\omega}{\omega})^2.
\end{equation}
This method overestimates the effects of random errors since it does not distinguish between contributions to the real and imaginary parts of the dispersion.
To determine the maximum tolerance for a given error, we consider each error acting alone.
The results of this calculation are summarized in Table \ref{tbl:tolerances}.
The dramatically worse tolerances for the polarizers in the diode are due to reliance on
resonant losses, which constrains $B(T)$.
However, the operation of the polarizers is perhaps the least important part of the diode.
So long as they produce appreciable losses, their precise magnitude is unimportant.
Hence we can more easily accept errors here than other parts of the circuit.
Moreover, we can always improve polarizers by increasing their thickness.

\begin{table}
\begin{tabular}{|c|c|c|c|}
\hline 
 & Polarizer & Gyrator & Transistor\tabularnewline
\hline
\hline 
$\delta H$ & 81.5$\mu$G & 3.42G & 25.0G\tabularnewline
\hline 
$\delta\omega$ & 3.32kHz & 34.2MHz & 49.5MHz\tabularnewline
\hline 
$\delta L$ & 30.0$\mu$m & 12.0$\mu$m & 10.0$\mu$m\tabularnewline
\hline
\end{tabular}
\caption{\label{tbl:tolerances}Maximum allowable tolerances for errors in independent parameters,
assuming 1\% operational error.
Each calculation assumes that other errors are 0.
The "polarizer" and "gyrator" columns refer to parts of the diode.
Tolerances for the transistor are calculated in the off state, which are more stringent.}
\end{table}

This trade-off between performance and thickness is a common feature in our circuits.
Ergo, it is worth considering some of the problems that might hinder circuit minimization.
Here, we considered systems with length scales in the mm-cm range because this possessed the most
robust body of experimental literature \cite{GG Book, TGG Exp, Magnons, Luthi Book, YIG Exp, EXP}.
However, for practical computers, working with smaller feature sizes is preferable.
This has several difficulties for our approach.
The most fundamental limit is that, for 10GHz phonons in YIG (as we consider here), the wavelength is
about 2.5$\mu$m.
For feature sizes smaller than a wavelength, the assumption that the device can simply be treated as a continuous
medium ceases to be applicable and we are forced to treat our devices as defects in a background medium.
To exceed this limit, would likely require even higher frequency phonons, where techniques to
prepare and measure shear waves are less developed \cite{Pico Phn, Pico Phn2}.
Even before we reach this limit, shrinking the system while maintaining the same effect (i.e. $k_{new}(L_{new})
=k_{old}(L_{old})$) is a non-trivial demand.
For gyrators, in the off-state limit ($B\to\infty$), the phase acquired is proportional to $LT^2/B^2$.
Since we don't care about decreasing the phase acquired, then we can simply allow $L$ to decrease without needing
to modify any other parameters.
In the on-state ($B\to0$), however, the phase is proportional to $LT^{3/2}$ (for small $T$).
Shrinking $L$ therefore requires a concomitant increase in $T$ (and only result in an approximate
invarience) or a modification of the material used.
Finally, for the polarizers, assuming that we're on-resonance ($B=B^*(T)$), then the requirement of phase invariance
is quite similar to the active gyrator (although not as strict, since a more effective polarizer is still acceptable).
To modify the MA material is therefore likely necessary for miniaturizing our circuits.
This could be done in several ways.
The most promising modifications of this approach would be to use single molecule magnets, which also show 
MA properties \cite{SMM}, or a bulk MA with a reduced speed of sound
(exposing the phonons to the MA for longer).

We have demonstrated the operation and limitations of phonon logic circuits outside of the electronic circuit paradigm.
Diodes and transistors remain difficult to construct for phonons, but the
MA approach presented here avoids many of the problems found
in other techniques.
While it faces challenges not present in previous approaches (e.g. minimization),
here we demonstrate that proof-of-concept realizations are feasible. 
We find that, not only are the requisite experimental conditions within an accessible range,
but also that such circuit elements should be sufficiently robust that noise should not effect them. 

This material is based upon work supported by the National Science Foundation Graduate Research Fellowship under Grant No. 1122374.


\begin{thebibliography}{25}

\bibitem{Phononics}N. Li, J. Ren, L. Wang, G. Zhang, P. H\"{a}nggi, and B. Li, \textit{Rev. Mod. Phys.} \textbf{84},
1045-1066 (2012).

\bibitem{Phn Logic}L. Wang and B. Li, \textit{Phys. Rev. Lett.} \textbf{99},
177208 (2007).

\bibitem{Phn Diode}B. Li, L. Wang, and G. Casati, \textit{Phys. Rev. Lett.} \textbf{93},
184301 (2004).

\bibitem{Phn Transistor}W.C. Lo, L. Wang, and B. Li, \textit{J. Phys. Soc. Japan}
\textbf{77}, 054402 (2008).

\bibitem{Phn Mem}L. Wang and B. Li, \textit{Phys. Rev. Lett.} \textbf{101},
267203 (2008).

\bibitem{CNT Rect}C. Chang, D. Okawa, A. Majumdar, and A. Zettl, \textit{Science} \textbf{314},
1121 (2006).

\bibitem{VO Rect}R.-G. Xie, C-T. Bui, B. Varghese, M.-G. Xia,  Q.-X. Zhang, C.-H. Sow,  B. Li, and J.T.L. Thong, \textit{Adv. Funct. Mater.} \textbf{21},
1602 (2011).

\bibitem{Therm Rect}M. Terraneo, M. Peyrard, and G. Casati, \textit{Phys. Rev. Lett.}
\textbf{88}, 094302 (2002).

\bibitem{Kittel} C. Kittel, \textit{Phys. Rev.} \textbf{110}, 836
(1958).

\bibitem{GG Book} J.D. Gavenda, V.V. Gudkov, \textit{Magnetoacoustic
Polarization Phenomena in Solids} (Springer, New York, NY, 2000).

\bibitem{AFM Th}M. Boiteux, P. Doussineau, B. Ferry, J. Joffrin, and A. Levelut, \textit{Phys. Rev. B} \textbf{4},
3077 (1971).

\bibitem{TGG Exp}A. Sytcheva, U. L\"{o}w, S. Yasin, J. Wosnitza, S. Zherlitsyn, P. Thalmeier, T. Goto, P. Wyder, and B. L\"{u}thi , \textit{Phys. Rev. B} \textbf{81},
214415 (2010).

\bibitem{Luthi Book} B. L\"{u}thi, \textit{Physical Acoustics in the
Solid State} (Springer, New York, NY, 2007).

\bibitem{Magnons}M. Bombeck \textit{et al.}, \textit{Phys. Rev. B} \textbf{85},
195324 (2012).

\bibitem{EXP}R. Guermeur, J. Joffrin, A. Levelut, and J. Penne, \textit{Solid State Commun.} \textbf{5}, 369 (1967)

\bibitem{Pico Phn}D. Hurley, O.B. Wright, O. Matsudaa, V.E. Gusevc, and O.V. Kolosov, \textit{Ultrasonics} \textbf{38},
470 (2000).

\bibitem{Pico Phn2}O. Matsuda, O. B. Wright, D. H. Hurley, V. E. Gusev, and K. Shimizu, \textit{Phys. Rev. Lett.}
\textbf{93}, 095501 (2004).

\bibitem{YIG Exp}H. Matthews and R.C. LeCraw, \textit{Phys. Rev.
Lett.} \textbf{8}, 397 (1962).

\bibitem{Pht Isolator}J. Ballato and E. Snitzer, \textit{Appl. Opt.}
\textbf{34}, 6848-6854 (1995). 

\bibitem{SMM}I. D. Tokman, V. I. Pozdnjakova, and A. I. Beludanova, \textit{Phys. Rev. B} \textbf{83}, 014405 (2011).

\end{thebibliography}
\end{document}